\documentclass[11pt,english,superscriptaddress,nofootinbib,preprintnumbers, aps]{revtex4}

\usepackage{amsmath,amsthm,latexsym,amssymb,amsfonts,epsfig}

\usepackage{float}
\usepackage{upgreek}

\usepackage{xcolor}
\usepackage{amsmath}
\usepackage{amsthm}
\usepackage{graphicx}
\usepackage{amssymb}
\usepackage{physics}
\usepackage{mathtools}
\usepackage{dsfont}
\usepackage{color}
\usepackage{soul}
\usepackage[hyperfootnotes=true,unicode=true, 
 bookmarks=true,bookmarksnumbered=false,bookmarksopen=false,
 breaklinks=false,pdfborder={0 0 1},backref=false,colorlinks=true]
 {hyperref}
 \usepackage{float}
 \usepackage[markup=nocolor]{changes}

\usepackage[makeroom]{cancel}

\usepackage{etoolbox}

\usepackage{hyperref}
\hypersetup{
    colorlinks,%
    citecolor=blue,%
    filecolor=blue,%
    linkcolor=blue,%
    urlcolor=blue
}

\newtheorem{Theorem}{Theorem}
\newtheorem{Definition}{Definition}
\newtheorem{Lemma}{Lemma}

\newcommand{\m}{\mbox{}}

\newcommand{\be}{\begin{equation}}
\newcommand{\ee}{\end{equation}}
\newcommand{\ba}{\begin{eqnarray}}
\newcommand{\ea}{\end{eqnarray}}
\def\bal#1\eal{\begin{align}#1\end{align}}

\makeatletter
\@addtoreset{equation}{section}
\makeatletter
\@ifundefined{textcolor}{}
{%
 \definecolor{BLACK}{gray}{0}
 \definecolor{WHITE}{gray}{1}
 \definecolor{RED}{rgb}{1,0,0}
 \definecolor{GREEN}{RGB}{0,204,0}
 \definecolor{BLUE}{rgb}{0,0,1}
 \definecolor{CYAN}{cmyk}{1,0,0,0}
 \definecolor{MAGENTA}{cmyk}{0,1,0,0}
 \definecolor{YELLOW}{cmyk}{0,0,1,0}
 }

\usepackage{hyperref}
\hypersetup{
    colorlinks,%
    citecolor=blue,%
    filecolor=blue,%
    linkcolor=blue,%
    urlcolor=blue
}

\makeatother

\newcommand{\re}{\mathbb{R}}
\newcommand{\Hil}{\mathcal{H}}
\newcommand{\id}{\mathds{1}}

\newcommand{\deta}{\partial_\eta}

\NewDocumentCommand\repTrace{m}{\Trace^{(#1)}}
\NewDocumentCommand\repTr{m}{\opbraces{\repTrace{#1}}}

\NewDocumentCommand\func{moo}{\opbraces{\mathop{#1\m}\nolimits\IfNoValueTF{#2}{}{^{#2}}\IfNoValueTF{#3}{}{_{#3}}}}
\NewDocumentCommand\fomega{o}{\func{\omega}[][#1]}
\NewDocumentCommand\fOmega{o}{\func{\Omega}[#1]}

\begin{document}

\title{Expectation values of Coherent States for ${\rm SU}(2)$ Lattice Gauge Theories}

\author{Klaus Liegener}
\email{liegener1@lsu.edu}
\affiliation{Department of Physics and Astronomy, Louisiana State University, Baton Rouge, LA 70803, USA}

\author{Ernst-Albrecht Zwicknagel}
\email{ernst-albrecht.zwicknagel@fau.de}
\affiliation{Institute for Quantum Gravity, Friedrich-Alexander University Erlangen-N\"urnberg, Staudtstra\ss e 7, 91058 Erlangen, Germany}

\date{\today{}}

\begin{abstract}

\fontfamily{lmss}\selectfont{
This article investigates properties of semiclassical Gauge Field Theory Coherent States for general quantum gauge theories. Useful, e.g., for the canonical formulation of Lattice Gauge Theories these states are labelled by a point in the classical phase space and constructed such that the expectation values of the canonical operators are sharply peaked on said phase space point. For the case of the non-abelian gauge group ${\rm SU}(2)$, we will explicitly compute the expectation value of general polynomials including the first order quantum corrections. This allows asking more precise questions about the quantum fluctuations of any given semiclassical system.}
\end{abstract}

\maketitle
\section{Introduction}
\label{s1}
\numberwithin{equation}{section}
One of the astonishing facts of modern physics is that many of its most powerful theories can be described using gauge symmetries. To understand their quantisation, a promising avenue comes in the study of Lattice Gauge Theories (LGT) \cite{Cre83,MM94,Gup98,MW00,Smi02,HLS17}. On the one hand, high performance computations in lattice quantum chromodynamics are the main tool to aid the experiments in particle and nuclear physics \cite{USQCD1,USQCD2}. On the other hand, LGT provides a theoretical framework that is ideally suited to make an impact on models trying to explore new physics beyond and within the standard model \cite{USQCD3,Sve18}. In the recent decades, methods from LGT have been further developed in the emergent field of quantum gravity, as it transpired that general relativity can be understood as an ${\rm SU}(2)$ gauge theory as well \cite{GP00,Rov04,Thi09}.\\
Further progress is much needed, since despite the active research on the quantisation of gauge theories not a single interacting 4-dimensional quantum Yang-Mills theory obeying the Wightman axioms has been constructed as of today. It remains one of the open millennium problems of the Clay Mathematical Institute \cite{cmi}. A possible route for attacking this caveat with the needed mathematical rigour might come in the Hamiltonian formulation to gauge theories. The latter has originally been developed by Kogut and Susskind for pure quantum Yang-Mills theories \cite{KS75} and enabled the construction of a well-defined kinematical Hilbert space, where the natural Haar measure on the compact gauge group can be used in order to define the Hilbert space measure. Nonetheless, implementing the dynamics of the theory poses a challenge: while a regularised Hamiltonian in presence of a finite ultraviolet cutoff is well defined, the necessary continuum limit is in general problematic. This caveat is hoped to be overcome in the renormalisation group program \cite{WK74,Wil75,Bal89a,Bal98b,Has98}, of which extensions to the Hamiltonian sector are currently under development \cite{Weg72,GW94,LLT}.\\
The present paper, however, will focus its analysis on the kinematical Hilbert space and the question of how \emph{semi-classical} field configurations can be recovered. In the canonical setting this is envisioned to be achieved by using so-called \emph{coherent states}. By this we mean states in the Hilbert space that are sharply peaked over classical data of the corresponding gauge theory for certain observables, e.g.\ the holonomies along edges of a lattice with fixed spacing $\epsilon$. Measurements on the classical system can then be thought of as the expectation values of the respective observables on these coherent states. However, the expectation value will differ from the classical value in the form of tiny quantum fluctuations, which should be measurable if the system is indeed described by the proposed underlying quantum theory.\\
Recently, several proposals for states to study semi-classical phenomena have emerged, e.g.\ Matrix Product states, useful for tensor network techniques \cite{VCM09,Oru14}, or Gaussian states for 1+1 ${\rm U}(1)$ and ${\rm SU}(2)$ LGT for variational studies \cite{SSKBDC18}. For arbitrary dimensions and any compact gauge group, a promising proposal for such semi-classical states came (extending preliminary work by Hall \cite{Hall94,Hall97}) from Thiemann et al. in form of the \emph{Gauge Field Theory Coherent States} (GCS) \cite{TW1,TW2,TW3,STW4}. Among the aforementioned peakedness property, these states saturate moreover the Heisenberg uncertainty bound for the fluctuations of connection and electric field, and present an overcomplete basis on the Hilbert space if certain analytical conditions are met. The present paper aims to use these GCS in the following way: We will fix a finite set of relevant observables that are given by a discrete lattice $\Gamma$ of spacing $\epsilon$ (and its dual cell complex), i.e.\ suitable smearings of connection and electric field. The full Hilbert space, which we are therefore interested in, can be decomposed as a tensor product over individual Hilbert spaces $\Hil_e$ for each edge $e\in\Gamma$. The GCS in $\Hil_e$ are square-integrable functions over the gauge group $G$, labelled by the holonomies and the gauge-covariant fluxes, i.e.\ smearings of the electric field, for the corresponding edge $e$ for some classical initial data. On the quantum level, one promotes holonomies to multiplication operators and fluxes to right-invariant vector fields. Due to the tensor product structure, computing the expectation value of any polynomial operator in holonomies and fluxes can be simplified: the operator can be split into multiple ones acting on each $\Hil_e$ separately and then the expectation values on each $\Hil_e$ can be computed independently. The aim of this paper is to give the concise formulas for the expectation values of general monomials of operators in any GCS on a single edge Hilbert space for the concrete case of gauge group ${\rm SU}(2)$. This will be done including linear order in the spread of the state, that means - if one chooses the spread of the state to be proportional to $\hbar$ - the tools presented enable the computation of any expectation value including the first order quantum corrections, thereby extending earlier work from \cite{DL17b}. Hence, the results are for example immediately applicable for calculations concerning electroweak interactions and theories such as Loop Quantum Gravity.\footnote{There also exist several results for the abelian group ${\rm U}(1)$ \cite{TW2,GT06}, and preliminary work for ${\rm SU}(3)$ \cite{LT16,Lan16}.}\\

The organisation of this article is as follows:\\
In section \ref{s2} we will repeat the construction of coherent states for gauge theories and their main features. For this purpose, it will be important to introduce a lattice discretisation of the classical solutions (or initial data) of connection and electric field. This comes in the form of holonomies along edges for the connection, which is a favoured choice as it transforms covariantly under gauge transformations. As it turns out, when making use of the dual lattice, a gauge-covariant discretisation of the electric field is possible as well. Once these data are chosen, one can peak the coherent states on them. We will then present the general formulas by which the expectation value of any polynomial operator in the GCS for each edge can be computed up to linear order in the spread of the state.\\
In section \ref{s3} and \ref{s4}, we supplement the proof for the aforementioned formulas. Using the results of earlier investigations \cite{DL17b}, it suffices to compute the expectation value for arbitrary representations of the holonomy operator. Since the calculation turns out to be quite lengthy, we will split it into four lemmas. Moreover, the computation will be restricted to linear order in the spread of the state, however necessary details are given by which these formulas could in principle be extended.\\
In section \ref{s5} we finish with conclusion and outlook for further research.

\section{Gauge Field Theory Coherent States for ${\rm SU}(2)$}
\label{s2}
\numberwithin{equation}{section}
In this section, we summarize the main concepts from canonical quantisation in Lattice Gauge Theories (LGT) and key results from the literature about Gauge Field Theory Coherent States (GCS).\\

Given a manifold $\mathcal{M}\cong \re \times \sigma$ on which a gauge theory shall be defined. Typically, $\sigma$ is chosen to be $\re^3$ or in general any $3$-dimensional manifold admitting a principal $G$-bundle with connection over $\sigma$. As a first step, we  introduce an infrared cutoff $R$ of $\sigma$, in the sense of working with the torus $\sigma_R=[0,R]^3\subset \sigma$ with periodic boundary conditions. In the following we will restrict our attention to the case of the compact gauge group $G={\rm SU}(2)$, as most tools for GCS have been constructed therewith. We will denote the Lie algebra valued connection as $A_a(x)=A_a^I(x)\tau_I$ with $\tau_I$ being a suitable basis of $\mathfrak{su}(2)$. The choice in the following is $\tau_I:=-i\sigma_I/2$, with $\sigma_I$ being the Pauli matrices, satisfying $\Tr(\tau_I\tau_J)=-\delta_{IJ}/2$. A different basis also used is the \emph{spherical basis} $\uptau_s$, $s\in\{-1,0,+1\}$, where $\uptau_{\pm}:=\pm (\tau_1\pm i\tau_2)/\sqrt{2}$ and $\uptau_0:=\tau_3$. These are subject to the algebra $[\uptau_+,\uptau_-]=i \uptau_0$, $[\uptau_{\pm},\uptau_0]=\pm i\uptau_{\pm}$.\\
The canonical conjugated momentum to $A^I_a(x)$ is the electric field $E^a_I(x)$, a Lie algebra valued vector density of weight 1. In other words, the elementary Poisson brackets read: ($I,J,...$ are internal indices of $\mathfrak{su}(2)$)
\begin{align}
\{E^a_I(x),E^b_J(y)\}=\{A_a^I(x),A_b^J(y)\}=0,\hspace{20pt}\{E^a_I(x),A^J_b(y)\}=\kappa_0 \delta^a_b \delta^J_I \delta^{(3)}(x,y)
\end{align}
with $\kappa_0$ being the coupling constant of the gauge theory.\\
Moreover, the phase space is subject to the Gauss constraint $G_J$:
\begin{align}\label{GaussConst}
 G_J=\partial_a E^a_J+\epsilon_{JKL} A^K_a E^a_L =0
\end{align}
with $\epsilon_{JKL}$ being the Levi-Civita symbol.\\
Proceeding as standard for LGT, the first step towards defining the quantum theory is introducing an ultraviolet cutoff of $\sigma$. This is done by introducing a cubic lattice $\Gamma$ with $\mathcal{N}$ points along each direction of the coordinates described by $x^a$. Let $R$ be the coordinate length of the torus with respect to the fiducial flat metric $\eta$ and denote by $\epsilon=R/\mathcal{N}$ the regulator of the discretisation, i.e.\ the coordinate length of each edge or link of the lattice.\\
Along said edges $e$ of the lattice, we will compute the holonomies $h(e)\in{\rm SU}(2)$ of the connection and along the associated faces $S_e$ of the dual cell complex \cite{ThiVII00} (whose intersection with the lattice we choose to be in the middle of each edge), we will compute the gauge-covariant fluxes $P(S_e):=P^J(S_e)\tau_J$. For an edge $e_k$ along direction $k\in\{\pm 1,\pm2,\pm3\}$ these read:
\begin{align}
h(e_k)&:=\mathcal{P}\exp(\int_0^1 \!\dd{t} A_a^J(e_k(t))\tau_J\dot{e}_k^a(t))\\
P^J(S_{e_k})&:=-2\Tr\biggl(\tau_J h\qty(e_{k,[0,1/2]})\biggl[\int_{S_{e_k}} \!\!\!\dd{x} h(\rho_x) {\ast E(x)} h(\rho_x)^\dagger \biggr] h\qty(e_{k,[0,1/2]})^\dagger \biggr)
\end{align}
where in the path ordered exponential the latest time values are ordered to the right, $\ast$ denotes the hodge star operator and $\rho_x$ is some choice of paths inside of $S_e$ such that $\rho_x(0)\in e_k$ and $\rho_x(1)=x$, whose details do not affect the gauge-covariance\footnote{Choosing the gauge-covariant fluxes $P$ instead of just the smeared electric field $E$ is motivated from the fact that they transform covariantly under gauge transformations, i.e.\ as $P(S_e)\mapsto g(e(0))P(S_e)g(e(0))^{-1}$. The standard fluxes do not have such a transformation property in the presence of a finite regularisation parameter $\epsilon$. For further details see \cite{ThiVII00}.}.\\
Consequently, all physical quantities can be approximated by functions of holonomies and gauge-covariant fluxes. E.g.\ to the continuum Yang-Mills Hamiltonian function one can assign a regularised expression which under quantisation agrees with the Kogut-Susskind Hamiltonian \cite{KS75}.\\

Geometric quantisation of this system is done by assigning to each edge $e$ a function in $\Hil_e=L_2({\rm SU}(2),\dd{\mu_H})$ with $\mu_H$ being the unique left- and right-invariant Haar measure over ${\rm SU}(2)$. The full Hilbert space of the whole lattice is then simply the tensor product over all square integrable functions on each edge, $\Hil_\Gamma:=\otimes_{e} \Hil_e$. The holonomies get promoted to bounded, unitary multiplication operators and the fluxes to essentially self-adjoint derivation operators.\footnote{For recent work choosing different formulations see e.g.\ \cite{AR14,MS15,Ray18}.} If we label in the position representation $f\in\Hil_\Gamma$ as $f=f(\{g_e\}_{e\in\Gamma})$ then:
\begin{align}
\hat{h}^{(k)}_{mn}(e')f(\{g_e\}_{e\in\Gamma})&:=D^{(k)}_{mn}(g_{e'})f(\{g_e\}_{e\in\Gamma})\\
\hat{P}^I(S_{e'})f(\{g_e\}_{e\in\Gamma})&:=i\hbar\kappa_0R^I(e')f(\{g_e\}_{e\in\Gamma})
\end{align}
with  $D^{(k)}_{mn}(g_e)$ being the Wigner-matrix of group element $g_e$ in the $(2k+1)$-dimensional irreducible representation of ${\rm SU}(2)$ and the \emph{right-invariant vector field} $R^I(e)$:
\begin{align}
R^I(e')f(\{g_e\}_{e\in\Gamma}):=\eval{\dv{s}}_{s=0}f(...,e^{s\tau_I}g_{e'},...)
\end{align}
which obey the following commutation relations: ($r\in\{-1,0,+1\}$, $s,s'\neq 0$)
\begin{align}\label{CommRel}
&[R^r(e),\hat{h}^{(k)}_{ab}(e')]=\delta_{ee'}[\uptau_r]^{(k)}_{ac} \,\hat{h}^{(k)}_{cb}(e),\\& [R^s(e),R^{s'}(e')]=-i\delta_{ee'}\frac{s-s'}{2} \,R^0(e),\hspace{20pt}[R^s(e),R^0(e')]=-i\delta_{ee'}s \,R^s(e)\nonumber
\end{align}
where $[\uptau_r]^{(k)}:=(R^r D^{(k)})(\id)$ is the induced Lie algebra representation.\\

We will now describe the form and properties of the Gauge Field Theory Coherent States. For their derivation, the reader is referred to the literature \cite{TW1,TW2,TW3}.\\
Based on the idea of approximating a given classical field configuration $(\tilde{A},\tilde{E})$, one computes first for each lattice edge $e$ the corresponding smeared quantities $(\tilde{h}(e), \tilde{P}(S_e))$ and maps this to a complex polarisation of the classical phase space, i.e.\ $(\tilde{h}(e), \tilde{P}(S_e))\mapsto h^\mathbb{C}_e\in {\rm SL}(2,\mathbb{C})$, that expresses the complex connection as a function of the real phase space. For example, the left-polar decomposition prescribes:
\begin{align}\label{lpDecomp}
h^{\mathbb{C}}_e:=\exp(-\frac{it}{\hbar \kappa_0} \tilde{P}^J(S_e)\tau_J)\tilde{h}(e)
\end{align}
The so called \emph{semiclassicality parameter} $t\geq 0$ is an arbitrary, dimensionless parameter (in typical applications, one often chooses $t=\hbar\kappa_0/\ell^2 $ where $\ell$ is some length scale).
\begin{Definition}[GCS]
The \emph{Gauge Field Theory Coherent State} $\psi_{h^\mathbb{C}}^t\in\Hil_e$ for each edge $e$ and classical field configuration $h^\mathbb{C} \in {\rm SL}(2,\mathbb{C})$ is given by 
\begin{align}\label{CohStaDef}
\psi^t_{h^\mathbb{C}}(g):=\sum_{j\in\mathbb{N}_0/2}d_j e^{-t(d_j^2-1)/8} \repTr{j}(h^\mathbb{C} g^\dagger)
\end{align}
where $t\geq 0$, $d_j=2j+1$ and $\repTr{j}(.)$ denotes the trace in the spin-$j$ irreducible representation of ${\rm SU}(2)$. Finally, $\expval{1}:=\norm*{\psi^t_{h^\mathbb{C}}}^2$ denotes the normalisation of the state.
\end{Definition}

As was shown in \cite{TW2,TW3}, these GCS are sharply peaked on the classical configuration in the following sense:
\begin{Theorem}
Let $\psi^t_{h^\mathbb{C}},\psi^t_{g^\mathbb{C}}\in\Hil_e$ be two GCS. For all $h^\mathbb{C},g^\mathbb{C}\in {\rm SL}(2,\mathbb{C})$ there exists a positive function $K_t(h^\mathbb{C},g^\mathbb{C})$ decaying exponentially fast as $t\to0$ for $h^\mathbb{C}\neq g^\mathbb{C}$ and such that
\begin{align}
\abs\big{\langle\psi^t_{h^\mathbb{C}},\psi^t_{g^\mathbb{C}}\rangle}^2\leq K_t(h^\mathbb{C},g^\mathbb{C})\norm\big{\psi^t_{h^\mathbb{C}}}^2 \norm\big{\psi^t_{g^\mathbb{C}}}^2
\end{align}
Moreover, for holonomy and flux operators on $\Hil_e$ one finds 
\begin{align}
\langle \psi^t_{h^\mathbb{C}},\hat{h}^{(\frac{1}{2})}(e)\psi^t_{g^\mathbb{C}}\rangle&=h(e)\langle\psi^t_{h^\mathbb{C}},\psi^t_{g^\mathbb{C}}\rangle+\order{t}\\
\langle \psi^t_{h^\mathbb{C}},\hat{P}^J(S_e)\psi^t_{g^\mathbb{C}}\rangle&=P^J(S_e)\langle \psi^t_{h^\mathbb{C}},\psi^t_{g^\mathbb{C}}\rangle+\order{t}
\end{align}
where $h(e)$ and $P^J(e)$ stem from the decomposition of $h^\mathbb{C}$ in (\ref{lpDecomp}).
\end{Theorem}

These statements were also extended to general polynomial operators in the basic configuration variables in \cite{TW3}. Hence, the GCS prove as useful tool for testing quantum systems which appear highly classical. For such a purpose it is however interesting to study the small corrections that appear for a finite semiclassicality parameter $t$, as these should be found when performing a measurement on such a semiclassical system.\\
First steps towards this task have been undergone in \cite{DL17b}. Based on the observation that any element $h\in {\rm SL}(2,\mathbb{C})$ may be written in its \emph{holomorphic decomposition} \cite{Car00,BMP09} as
\begin{align}
h=n\,e^{-(\xi-i\eta)\tau_3}\,\tilde{n},\hspace{30pt}\xi,\eta\in\mathbb{R},\hspace{10pt} n,\tilde{n}\in {\rm SU}(2)
\end{align}
formulas for the expectation values of some polynomial operators were found. However, the investigations in \cite{DL17b} were restricted to operators involving a single holonomy operator $\hat{h}^{(k)}$ with $k=1/2,1$. But since the matrix elements of any product of representations of ${\rm SU}(2)$ can be expressed as a sum over the matrix elements of higher irreducible representations (due to the Peter \& Weyl theorem \cite{PW27}), having knowledge of $\langle \psi^t_h, \hat{h}^{(k)} \psi^t_h\rangle$ for any $k\in \mathbb N/2$ is the last missing piece to compute the expectation value of any polynomial observable in holonomies and right-invariant vector fields neglecting $\order*{t^2}$ corrections. This article fills that gap by proving the following theorem:
\begin{Theorem}\label{TheoremResult}
Let $\psi^t_h\in\Hil_e$ be a GCS. Then the expectation value of a holonomy operator on edge $e$ is given by ($\expval{.} := \langle \psi^t_h,\,.\;\psi^t_h\rangle$)
\begin{align}\label{Result}
\frac{\langle \hat{h}^{(k)}_{ab}\rangle}{\expval{1}} = D^{(k)}_{ac}(n) e^{i\xi c} \gamma^k_c D^{(k)}_{cb}(\tilde{n})
\end{align}
with
\begin{align}\label{GammaResult}
\gamma^k_a&=1-\frac{t}{4}\qty[\qty(k(k+1)-a^2)\frac{\tanh(\eta/2)}{\eta/2}+a^2]+\order*{t^2}
\end{align}
and
\begin{align}
\label{normalisation}
\expval{1} :\mskip-\thickmuskip&=\sqrt{\frac{\pi}{t^3}}\frac{2\eta e^{\eta^2/t}}{\sinh(\eta)}e^{t/4}
\end{align}
\end{Theorem}
The remaining sections of this article will substantiate the proof of this formula.\\

Equipped with this knowledge, we can also generalize the expectation values of monomials involving holonomies and right-invariant vector fields from \cite{DL17b} to the following statements:\footnote{Due to a different choice in conventions, in the formula from \cite{DL17b} we had to substitute $z\to -\bar{z}$ and divide by $2^{N}$.}
\begin{align}
\expval*{\hat{h}^{(k)}_{ab}R^{K_1}...R^{K_N}}
&=\expval*{\hat{h}^{(k)}_{cb}}
\qty(\frac{i\eta}{t})^N  D^{(1)}_{-K_1-S_1}(n)...D^{(1)}_{-K_N-S_N}(n) \label{hRRR}\\
&\hspace{15pt}\times\Biggl(\delta^{S_1}_0...\delta^{S_N}_0\delta_{ac}
+\frac{t}{2\eta}\Biggl[\delta^{S_1}_0...\delta^{S_N}_0\delta_{ac} \,N\qty(\frac{N+1}{2\eta}-\coth(\eta)) \nonumber\\
&\hspace{30pt}+i\sum_{A=1}^N\delta^{S_1}_0...\cancel \delta^{S_A}_0 ...\delta^{S_N}_0 \qty\big(1-S_A\tanh(\eta/2)) D^{(1)}_{-S_A-L}(n^\dagger)[\uptau_L]^{(k)}_{ac} \nonumber\\
&\hspace{30pt}-\frac{\delta_{ac}}{\sinh(\eta)}\sum_{A<B=1}^N\delta^{S_1}_0...\cancel \delta^{S_A}_0...\cancel \delta^{S_B}_0... \delta^{S_N}_0(\delta^{S_A}_{+1}\delta^{S_B}_{-1}+\delta^{S_A}_{-1}\delta^{S_B}_{+1})e^{S_A\eta}\Biggr]+\order*{t^2}
\Biggr)\nonumber
\end{align}
Take note that this formula includes the case $k=0$ which corresponds to no appearing holonomy operators. We stress again that any monomial in holonomies and right-invariant vector fields can always be brought into the form (\ref{hRRR}) by suitably using the commutator relations (\ref{CommRel}) and ${\rm SU}(2)$ recoupling theory. Also, in order to extend these formulas to left-invariant vector fields we refer to section III.B of \cite{DL17b}.\\

Finally, let us comment on the implementation of the quantum Gauss constraint. A tensor product consisting solely of GCS on each edge will in general not be a solution to the standard quantisation of (\ref{GaussConst}). To satisfy the Gauss constraint at the quantum level there are several mechanisms. As is customary in LGT, one could choose a maximal tree graph $\Gamma_T\subset\Gamma$ without closed loops on which the gauge is fixed \cite{Cre83,Cre77,DKLL85,Ada02}. Another procedure, which became popularized in Quantum Gravity approaches on the lattice, is the so called group averaging procedure \cite{Thi09,ALMMT95,Mar95,Mar99,Mar00}, by which e.g.\ a simple tensor product over coherent states can be projected to a gauge invariant state \cite{Bahr}.\footnote{It is worthwhile to note that the expectation value of gauge-invariant operators, like Wilson loops, does not get affected by this projection in its leading order in $t\sim \hbar$.}

\section{Outlining the proof of Theorem \ref{TheoremResult}}
\label{s3}
\begin{Lemma}
\label{Lemma_Gamma}
From the definition of the coherent states, it follows straightforward (\ref{Result}). Using Wigner $3j$-symbols to express the action of the holonomy operator, one finds explicitly for the coefficient $\gamma$:
\begin{align}
\gamma^J_M=\frac{1}{\expval{1}}\sum_{M'=-J}^J\beta(M')\!\!\! \sum_{j\geq (J-M')/2}\sum_{m=-j}^je^{-t(d_j+M')^2/4}\frac{d_j(d_j+2M')}{Q(d_j+M')}e^{2m\eta}\fomega(j,m)\fOmega[2](j,m)\label{gamma_original}
\end{align}
with the following definitions:
\begin{align*}
\Delta^\pm&:=J\pm\max(|M|,|M'|),\hspace{80pt}\delta^\pm:=J\pm(|M'+M|-|M'-M|)/2\\
\beta(M')&:=2^{-2J}e^{t(1-{M'}^2)/4}e^{\eta M} \frac{\Delta^+!\Delta^-!}{\delta^+!\delta^-!},\hspace{90pt}n^\pm:=-\min(0,M'\pm M)\\
\fOmega(j,m)&:=2^{J}\sum_{k=0}^{\Delta^-}(-)^{k}\binom{\delta^+}{k}
\binom{\delta^-}{\Delta^--k}
\fomega[k](j, m),\hspace{30pt}Q(v):=\prod_{N=-J}^J(v-N)\\
\fomega(j,m)&:=\frac{(j+m+\max(0,M'+M))!}{(j+m+\min(0,M'+M))!}\frac{(j-m+\max(0,M'-M))!}{(j-m+\min(0,M'-M))!}\\
\fomega[k](j,m)&:= \frac{(j-m-n^-)!}{(j-m-k-n^-)!}\frac{(j+m-n^+)!}{(j+m+k-\Delta^--n^+)!}
\end{align*}
\end{Lemma}
For the following statements we will assume that $\eta\neq 0$ and later extend them by taking the limit, realizing that it does not change the computed corrections.
\begin{Lemma}
\label{Lemma_Extension}
It is possible to rewrite (\ref{gamma_original}) such that $\gamma^J_M=\tilde{\gamma}^J_M/\expval{1}-\order{t^\infty}$ and
\begin{align}
\tilde{\gamma}^J_M=\sum_{M'=-J}^J\frac{\beta}{2}\sum_{u\in\mathbb{Z}}e^{-t(u+M')^2/4}\frac{u(u+2M')}{Q(u+M')}\fomega(\frac{u-1}{2},\frac{\deta}{2})\fOmega[2](\frac{u-1}{2},\frac{\deta}{2})\frac{\sinh(u\eta)}{\sinh(\eta)}
\end{align}
\end{Lemma}
At this point one can use the elementary Poisson summation formula, i.e.\ \cite{Bochner}
\begin{Theorem}[Poisson Summation Formula]
Consider $f\in L_1(\mathbb{R},dx)$ such that the series $\sum_{n\in\mathbb{Z}}f(y+ns)$ is absolutely and uniformly convergent for $y\in [0,s],\; s>0$. Then
\begin{align}
\sum_{n\in\mathbb{Z}}f(ns)=\sum_{n\in\mathbb{Z}}\int_{\mathbb{R}} \dd{x} e^{-2\pi inx}f(sx)
\end{align}
\end{Theorem}
\begin{Lemma}\label{Lemma_Coefficient}
After applying the Poisson Summation Formula, one can neglect all terms of the outer sum which are of order $\order{t^\infty}$, i.e.\ all terms except $n=0$. Then
\begin{align}\label{insertedHeaviside}
\gamma^J_M=\frac{1}{\expval{1}}\sum_{M'=-J}^J\frac{\beta}{2}\int_{\mathbb{R}}\dd{v} e^{-tv^2/4+v\eta}\frac{\chi(v)}{Q(v)} \qty[\sum_{k=0}^{d_J+1} P_k v^k]
\end{align}
where $\chi = 1$ everywhere but on a compact set, and especially the two highest coefficients of the polynomial $P$ are given by:
\begin{align}
P_{d_J+1}&=\frac{2^{2J}e^{-M\eta}}{\sinh(\eta)}\delta_{M,M'}\\
P_{d_J}&=\frac{2^{2J}e^{-M\eta}}{\sinh(\eta)}\qty[-\qty(J(J+1)-M^2)\coth(\eta)\,\delta_{M,M'}+\frac{(\delta^-)^2}{2\sinh(\eta)}\,\delta_{M\pm1,M'}]
\end{align}
\end{Lemma}

\begin{Lemma}\label{Lemma_Expansion}
The integral can be expanded including the linear order in $t$ as
\begin{align}\label{Lemma_Expansion_formula}
\gamma^J_M=\sum_{M'=-J}^J\frac{e^{\eta M} \sinh(\eta)\Delta^+!\Delta^-!}{2^{2J+2}\eta\;\delta^+!\delta^-!}\left[
4\eta {P}_{d_J+1}+(2P_{d_J}-\eta {M'}^2{P}_{d_J+1})t+\order*{t^2}
\right]
\end{align}
and upon inserting ${P}_k$ from the previous lemma explicitly the final result is:
\begin{align}\label{final_result_sec3}
\gamma^J_M = 1-\frac{t}{4}\qty[\qty(J(J+1)-M^2)\frac{\tanh(\eta/2)}{\eta/2}+M^2]+\order*{t^2}
\end{align}
\end{Lemma}
Lastly, we can perform trivially the limit of $\eta\to 0$ on the right side of (\ref{final_result_sec3}) which must agree with taking the limit in the expectation value of $\expval*{\hat{h}^{(k)}}$ due to strong continuity.

\section{Explicit proof of Theorem \ref{TheoremResult}}
\label{s4}

\subsection{Proof of Lemma \ref{Lemma_Gamma}}
Using the standard recoupling techniques for ${\rm SU}(2)$, i.e.\ \cite{BS68}
\begin{align}\label{SU(2)Mulitplication}
D^{(j_1)}_{ab}(g)D^{(j_2)}_{cd}(g)=\sum_{j=|j_1-j_2|}^{j_1+j_2}d_j (-)^{m-n}\left(\begin{array}{ccc}
j_1 & j_2 & j\\
a & c & m
\end{array}\right)\left(\begin{array}{ccc}
j_1 & j_2 & j\\
b & d & n
\end{array}\right)D^{(j)}_{-m-n}(g)
\end{align}
one obtains easily - starting from (\ref{CohStaDef}) with $h\equiv h^\mathbb{C}\in {\rm SL}(2,\mathbb{C})$ - that:
\begin{align}
\frac{\langle \hat{h}^{(k)}_{ab}\rangle}{\expval{1}}&=\frac{1}{\expval{1}}\sum_{j,j'\geq 0}d_jd_{j'} e^{-t(d_j^2+d^2_{j'}-2)/8}\int_{{\rm SU}(2)} \!\!\dd{\mu_H(g)} \overline{\repTrace{j}(h g^\dagger)}D^{(k)}_{ab}(g)\repTrace{j'}(h g^\dagger)\nonumber\\
&=\frac{1}{\expval{1}}\sum_{j,j'\geq 0}d_{j}d_{j'}e^{-t(d_j^2+d_{j'}^2-2)/8}\int_{{\rm SU}(2)} \!\!\dd{\mu_H(g)} \repTrace{j}(e^{(\xi+i\eta)\tau_3} g)D^{(k)}_{ab}(ng\tilde n)\repTrace{j'}(e^{-(\xi-i\eta)\tau_3} g^\dagger)\nonumber\\
&=\frac{1}{\expval{1}}\sum_{j,j'\geq 0}d_{j}d_{j'}e^{-t(d_j^2+d_{j'}^2-2)/8+i(\xi-i\eta)c'-i(\xi+i\eta)c}D^{(k)}_{aa'}(n)D^{(k)}_{b'b}(\tilde n)\int_{{\rm SU}(2)} \!\!\dd{\mu_H(g)}\times\nonumber\\
&\hspace{50pt}\times\sum_{J=|k-j|}^{k+j} d_J(-)^{m-n}\left(\begin{array}{ccc}
	k & j & J\\
	a' & c & m
\end{array}\right)\left(\begin{array}{ccc}
	k & j & J\\
	b' & c & n
\end{array}\right) D_{c'c'}^{(j')}(g^\dagger) D^{(J)}_{-m-n}(g)\nonumber\\
&=\frac{1}{\expval{1}}\sum_{j,j'\geq 0}d_{j}d_{j'}e^{-t(d_j^2+d_{j'}^2-2)/8+i\xi(c'-c)+\eta(c+c')}D^{(k)}_{aa'}(n)D^{(k)}_{b'b}(\tilde n)
\left(\begin{array}{ccc}
	k & j & j'\\
	a' & c & -c'
\end{array}\right)\left(\begin{array}{ccc}
	k & j & j'\\
	b' & c & -c'
\end{array}\right)\nonumber\\
&=D_{aa'}^{(k)}(n)\delta_{a'b'}e^{i\xi a'}D_{b'b}^{(k)}(\tilde n) \gamma^k_{a'}
\end{align}
where we used for the second line the left- and right-invariance of the Haar measure for $n^\dagger g\tilde n^\dagger\mapsto g$, for the third  (\ref{SU(2)Mulitplication}) and $D^{(k)}_{ab}(e^{z\tau_3})=\delta_{ab}e^{-iza}$, for the fourth orthogonality of the Wigner-$D$ functions and finally in the last line that $a'=c'-c=b'$ and the definition
\begin{align}
\gamma^J_{M}=\frac{1}{\expval{1}}\sum_{j,j'\geq 0}\sum_{m=-j}^j\sum_{m'=-j'}^{j'}d_jd_{j'}e^{-t(d_j^2+d_{j'}^2-2)/8}e^{\eta(m+m')}
\left(\begin{array}{ccc}
J & j & j'\\
M & m & -m'
\end{array}
\right)^2
\end{align}
as was claimed in (\ref{Result}). Due to the symmetry properties of the Wigner $3j$-symbol, it follows that:
\begin{align}
\gamma^J_M(\eta)=\gamma^J_M(-\eta)=\gamma^J_{-M}(\eta)
\end{align}
The coefficient $\gamma^J_M$ can now be further manipulated: The $3j$-symbol is defined to vanish unless $|j-J|<j'<j+J$, which allows us to replace the sum over $j'=j+M'$ by a corresponding sum over $M'$. Further it vanishes unless $m-m'+M=0$, which consumes the sum over $m'$:
\begin{align}\label{gamma3j}
\gamma^J_M=\frac{1}{\expval{1}}\sum_{M'=-J}^J\sum_{j\geq (J-M')/2}\sum_{m=-j}^j d_jd_{j+M'} e^{-t(d_j^2+d_{j+M'}^2-2)/8}e^{\eta(2m+M)}
\left(\begin{array}{ccc}
j & j+M' & J\\
m & -m-M & M
\end{array}
\right)^2
\end{align}
Note that we have truncated the sum to $j\geq (J-M')/2$ as the $3j$-symbol is zero for smaller $j$. Using that it vanishes also if $|m+M|>j+M'$ and otherwise applying the Racah formula \cite{Rac42,Varshalovich} gives rise to the following expression, which has no poles for the specified range of $j$:
\begin{align}\label{Racah}
&\left(\begin{array}{ccc}
j & j+M' & J\\
m & -m-M & M
\end{array}
\right)^2=(J+M)!(J-M)!(J+M')!(J-M')!\left(\prod_{N=-J}^J (d_j+M'-N)\right)^{-1}\\
&\;\;\;\;\;\times (j+m)!(j-m)!(j+m+M+M')!(j-m-M+M')!\nonumber\\
&\;\;\;\;\;\times \left( \sum_k (-)^k[k! (M'-M+k)! (J+M-k)! (J-M'-k)!(j-m-k)!(j+m+M'-J+k)!]^{-1}\right)^2\nonumber
\end{align}
where the sum runs over all $k\in\mathbb{N}_0$ such that the arguments of the factorials are non-negative.\\
Upon introducing the quantities
\begin{align}
\Delta^\pm:=J\pm \max(|M|,|M'|),\hspace{30pt}\delta^\pm:=J\pm(|M'+M|-|M'-M|)/2
\end{align}
which satisfy $\Delta^+\geq\delta^\pm\geq\Delta^-$,
we find
\begin{align}\label{firstDeltas}
(J+M)! (J-M)! (J+M')!(J-M')!=\Delta^+!\Delta^-!\delta^+!\delta^-!
\end{align}
and under a shift of the summation parameter $k\mapsto k+n^-$, with $n^\pm:=-\min(0,M'\pm M)$, we see:
\begin{align}
k!(M'-M+k)!\hspace{10pt}&\mapsto\hspace{10pt} k!(\delta^--\Delta^-+k)!,\\
(J+M-k)!(J-M'-k)!\hspace{10pt}&\mapsto\hspace{10pt} (\delta^+-k)!(\Delta^--k)!
\end{align}
Therefore, we obtain for the sum appearing in (\ref{Racah}):
\begin{align}\label{SumWithBinomial}
\left(\sum_k(-)^k \ldots\right)=\sum_k\frac{(-)^k}{\delta^+!\delta^-!}\binom{\delta^+}{k}
\binom{\delta^-}{\Delta^--k}
\frac{1}{(j-m-k-n^-)!(j+m+k-\Delta^--n^+)!}
\end{align}
Finally, the second line of (\ref{Racah}) can be rewritten via:
\begin{align}\label{split_kl_Dependence}
(j\pm m)! (j\pm m + M'\pm M)!= \frac{(j\pm m +\max(0,M'\pm M)!)}{(j\pm m +\min(0,M'\pm M)!)}\qty[(j\pm m - n^{\pm})!]^2
\end{align}
Plugging (\ref{firstDeltas}), (\ref{SumWithBinomial}) and (\ref{split_kl_Dependence}) into (\ref{Racah}) and the total result into (\ref{gamma3j}) gives:
\begin{align}
\gamma^J_M=\frac{1}{\expval{1}}\sum_{M'=-J}^J\beta(M')\!\!\! \sum_{j\geq (J-M')/2}\sum_{m=-j}^je^{-t(d_j+M')^2/4}\frac{d_j(d_j+2M')}{Q(d_j+M')}e^{2m\eta}\fomega(j,m)\fOmega[2](j,m)
\end{align}
where we have defined:
\begin{align}
\beta(M')&:=2^{-2J}e^{t(1-{M'}^2)/4}e^{\eta M} \frac{\Delta^+!\Delta^-!}{\delta^+!\delta^-!},\hspace{30pt}Q(v):=\prod_{N=-J}^J(v-N)\\
\fOmega(j,m)&:=2^{J}\sum_{k=0}^{\Delta^-}(-)^{k}\binom{\delta^+}{k}
\binom{\delta^-}{\Delta^--k}
\fomega[k](j, m)\\
\fomega(j,m)&:=\frac{(j+m+\max(0,M'+M))!}{(j+m+\min(0,M'+M))!}\frac{(j-m+\max(0,M'-M))!}{(j-m+\min(0,M'-M))!}\label{omega}\\
\fomega[k](j,m)&:= \frac{(j-m-n^-)!}{(j-m-k-n^-)!}\frac{(j+m-n^+)!}{(j+m+k-\Delta^--n^+)!}\label{omegak}
\end{align}
Note that the summation variables $m$ and $k$ are still subject to some implicit restrictions stemming from the Racah formula. This finishes the proof of lemma \ref{Lemma_Gamma}.

\subsection{Proof of Lemma \ref{Lemma_Extension}}
Upon closer inspection of (\ref{omega}) and (\ref{omegak}), we see that $\omega(j,m)$ and $\omega_k(j,m)$ are such that all factorials in the denominator cancel, resulting in polynomial expressions in $j\pm m$. In the following, these polynomials are therefore understood as the definition of $\omega$ and $\omega_k$, respectively, for all $j\in\mathbb{R}$.\\
Consider $\omega(j, m)\omega_k(j, m)$ as a polynomial in $m$, then some of its roots are in integer steps given by
\begin{align}
(j-k-n^-+1),\,...\,,(j)\hspace{50pt}(-j),\,...\,,(-j+\Delta^--k+n^+-1)
\end{align}
Note that $J-M' = \Delta^- +n^++n^-$ and thus for $j\leq(J-M'-1)/2$ we have:
\begin{align}
(j-n^-+1)=(2j-j-n^-+1)\leq (J-M'-j-n^-)=(-j+\Delta^-+n^+)
\end{align}
For any $\mathbb{N}_0/2 \ni j < (J-M')/2$ we can conclude that all $m$ from $-j$ to $j$ are roots of $\omega\omega_k$ for all $k$, and therefore also roots of $\omega\Omega$.
This allows us to extend the sum over $j\geq (J-M')/2$ to $j\geq 0$ by adding the corresponding counter terms, which are finite due to each pole in $1/Q$ being cancelled by a root of $\omega\Omega$. Following a similar argumentation we find that the implicit restrictions of $m$ and $k$ may be dropped thanks to the respective summands being zero. Hence,
\begin{align}
\gamma^J_M=\frac{1}{\expval{1}}
\tilde \gamma^J_M -R^J_M,\hspace{15pt}\tilde \gamma^J_M=\sum_{M'=-J}^J \sum_{j\geq 0}S(d_j,M'),\hspace{15pt}R^J_M=\frac{1}{\expval{1}}\sum_{M'=-J}^J \!\!\sum_{j=0}^{(J-M'-1)/2}\!\! S(d_j,M')
\end{align}
i.e.\ $R^J_M$ contains the counter terms, and with
\begin{align}
S(d_j,M'):=\beta(M') \,e^{-t(d_j+M')^2/4}\frac{d_j(d_j+2M')}{Q(d_j+M')}\sum_{m=-j}^je^{2m\eta}\fomega(j,m)\fOmega[2](j,m)
\end{align}
The normalisation of the state has been computed in \cite{DL17b} and reads:
\begin{align}
\expval{1} = \sqrt{\frac{\pi}{t^3}}\frac{2\eta e^{\eta^2/t}}{\sinh(\eta)}e^{t/4}
\end{align}
Therefore it is easy to estimate that
\begin{align}
\abs{R^J_M} \leq \sqrt{t^3}e^{-\eta^2/t}\sum_{M'=-J}^J \!\!\sum_{j=0}^{(J-M'-1)/2}\!\! \sum_{m=-j}^j
\abs{\beta|_{t=0} \frac{d_j(d_j+2M')}{Q(d_j+M')}e^{2m\eta}\fomega(j,m)\fOmega[2](j,m)\frac{\sinh(\eta)}{2\sqrt{\pi}\eta}}= C \sqrt{t^3} e^{-\eta^2/t}\nonumber
\end{align}
with $C$ being some finite constant independent of $t$. Since we are assuming $\eta\neq 0$, we see that $R^J_M=\order{t^\infty}$. It will hence be neglected in the following.\\

\noindent Before we continue: later on it will turn out to be useful to know that
\begin{align}\label{symmetrySummand}
S(u,M') = S(-u,-M')
\end{align}
In order to prove (\ref{symmetrySummand}), first note that
$\beta(M')=\beta(-M')$ and $Q(v)=(-)^{d_J}Q(-v)$.
From
\begin{align}\label{omega_Extended}
\fomega(j,m)\equiv \fomega(d_j,m,M') :=\!\prod_{i=\min(0,M'+M)+1/2}^{\max(0,M'+M)-1/2}\qty(\frac{d_j}{2}+m+i)\prod_{i=\min(0,M'-M)+1/2}^{\max(0,M'-M)-1/2}\qty(\frac{d_j}{2}-m+i)
\end{align}
it follows with $\max(0,-M'\pm M)=-\min(0,M'\mp M)$ that
\begin{align}
\fomega(-u,m,-M')=(-)^{2J}\fomega(u,m,M')
\end{align}
where we used that $2J-(|M'+M|+|M'-M|)=2\Delta^-$ is even.\\
For proving the respective symmetry of $\Omega(d_j,m,M'):=\Omega(j,m)$, we express $\Omega$ in terms of the \emph{generalized hypergeometric function} $_3{\rm F}_2$ (see \cite{Sla66} for properties thereof)
\begin{align}
\fOmega(d_j,m,M')&=2^J\binom{\delta^-}{\Delta^-}\frac{\Gamma(e-a)}{\Gamma(e)}\;_3{\rm F}_2(a,b,c;d,e;1) , \label{Omega1} \\
\fOmega(-d_j,m,-M')&=2^J\binom{\delta^-}{\Delta^-}\frac{\Gamma(e-a-b)}{\Gamma(e-b)}\;_3{\rm F}_2(a,b,d-c;d,a+b-e+1;1) \label{Omega2}
\end{align}
with the definitions:
\begin{align*}
a:=-\Delta^-,\hspace{12pt}b:=-\delta^+,\hspace{12pt}c:=-j+m+n^-,\hspace{12pt}d:=\delta^--\Delta^-+1,\hspace{12pt}e:=j+m+1-\Delta^--n^+
\end{align*}
Note that (\ref{Omega2}) is not equivalent to (\ref{Omega1}), but obtained by rewriting $\Omega(-d_j,m,-M')$ in its polynomial form first.
We now utilize the following transformation formula \cite{Bai72}:
\begin{align}\label{HyperGeoSymemtry}
\;_3{\rm F}_2&(a,b,c;d,e;1)= \frac{\Gamma(e)\Gamma(e-a-b)}{\Gamma(e-a)\Gamma(e-b)}\;_3{\rm F}_2(a,b,d-c;d,a+b-e+1;1)+\frac{\Gamma(d)\Gamma(e)\Gamma(a+b-e)}{\Gamma(a)\Gamma(b)\Gamma(d-c)}\times\nonumber\\
&\times\frac{\Gamma(d+e-a-b-c)}{\Gamma(d+e-a-b)}\;_3{\rm F}_2(e-a,e-b,d+e-a-b-c;d+e-a-b,e-a-b+1;1)
\end{align}
For $a,b\in\mathbb{Z}$\textbackslash$\mathbb{N}$ the first two functions $_3F_2$ are finite, and $1/(\Gamma(a)\Gamma(b))=0$. The third function certainly converges for $c-e> -1$, and (\ref{HyperGeoSymemtry}) then simplifies to
\begin{align}
_3F_2(a,b,c;d,e;1)=\frac{\Gamma(e)\Gamma(e-a-b)}{\Gamma(e-a)\Gamma(e-b)}\;_3F_2(a,b,d-c;d,a+b-e+1;1)
\end{align}
which tells immediately that (\ref{Omega1}) and (\ref{Omega2}) are the same expression if $c-e> -1$ or, equivalently, $d_j<J-M'+1$. However, since $\Omega(d_j,m,M')$ is polynomial in $d_j$ and $m$, this is already sufficient to conclude that $\Omega(u,m,M')=\Omega(-u,m,-M')$ everywhere.\\
Finally, due to the fact that $\omega(j,m)\Omega^2(j,m) $ is polynomial in $m$, we can use that
\begin{align}
\sum_{m=-j}^je^{2m\eta}m^k=\left(\frac{\deta}{2}\right)^k\frac{\sinh(d_j\eta)}{\sinh(\eta)}
\end{align}
(which is easily seen via performing a geometric sum) for the following replacement in the polynomial:
\begin{align}
\sum_{m=-j}^je^{2m\eta}\fomega(j,m)\fOmega[2](j,m)=\fomega(j,\deta/2)\fOmega[2](j,\deta/2)\frac{\sinh(d_j\eta)}{\sinh(\eta)}
\end{align}

Putting this and all the symmetries established for $\beta,Q,\omega,\Omega$ together, it follows (\ref{symmetrySummand}), as was claimed!\\

\noindent Having $S(u,M')=S(-u,-M')$ and $S(0,M')=0$ since $u\sinh(u\eta)/Q(u+M')|_{u=0}=0$ (and similar for multiple actions of $\deta$ thereon), we can write:
\begin{align}\label{extendend_sum}
\tilde{\gamma}^J_M&=\sum_{M'=-J}^J \sum_{j\geq 0} S(2j+1,M')=\sum_{M'=-J}^J \sum_{u=1}^\infty S(u,M')=\frac{1}{2}\sum_{M'=-J}^J\sum_{u\in\mathbb Z} S(u,M')
\end{align}
which finishes the proof of lemma \ref{Lemma_Extension}.

\subsection{Proof of Lemma \ref{Lemma_Coefficient}}

We apply the Poisson summation formula to (\ref{extendend_sum}):
\begin{align}\label{poissonformula_application}
\tilde{\gamma}^J_M
&= \frac{1}{2}\sum_{M'=-J}^J\sum_{n\in\mathbb{Z}}\int_{\mathbb{R}}\dd{u} e^{-2\pi i n u} S(u,M')\\
&=\sum_{M'=-J}^J\frac{\beta}{2}\sum_{n\in\mathbb{Z}}(-1)^{2M'n}e^{-4\pi^2n^2/t}\int_{\mathbb{R}}\dd{v}e^{-tv^2/4}\frac{u(u+2M')}{Q(u+M')}\fomega(\frac{u-1}{2},\frac{\deta}{2})\fOmega[2](\frac{u-1}{2},\frac{\deta}{2})\frac{\sinh(u\eta)}{\sinh(\eta)}\nonumber\\
&=\sum_{M'=-J}^J\frac{\beta}{2}\int_{\mathbb{R}}\dd{v} e^{-tv^2/4}\frac{v^2-{M'}^2}{Q(v)}\fomega(\frac{v-M'-1}{2},\frac{\deta}{2})\fOmega[2](\frac{v-M'-1}{2},\frac{\deta}{2})\frac{\sinh((v-M')\eta)}{\sinh(\eta)}\nonumber
\end{align}
where we introduced the new integration variable $v=u+M'+\frac{4\pi i n}{t}$ and in the last step used that $\exponential(-4\pi^2n^2/t)=\order{t^\infty}$ unless $n=0$.\\
Since $S(d_j,M')$ has no poles, it is clear that the integrand in the above formula has no poles as well. However, this is only true for this exact expression, which severely limits the possibilities of rewriting it. This motivates the following definition: Let $\chi(v)$ be a smooth, real, symmetric function satisfying $\chi(v)=0$ for $|v|\leq a$ and $\chi(v)=1$ for $|v|\geq b$ with $0<a<b$ fixed. By choosing $a>J$, the quotient $\chi/Q$ becomes a smooth function everywhere. Now, we split (\ref{poissonformula_application}) in two parts:
\begin{align}
\gamma^J_M =\frac{1}{\expval{1}} \tilde \gamma^J_M &=
\frac{1}{\expval{1}}\sum_{M'=-J}^J\frac{\beta}{2}\int_{\mathbb{R}}\dd{v} e^{-tv^2/4}\frac{\chi}{Q}\qty\big(v^2-{M'}^2)\,\omega\;\Omega^2\;\frac{\sinh((v-M')\eta)}{\sinh(\eta)}\\
&\quad+\qty(\sqrt{t^3}e^{-t/4}\frac{\sinh(\eta)}{2\eta\sqrt{\pi}}\frac{\beta}{2})e^{-\eta^2/t}\int_{\mathbb{R}}\dd{v} e^{-tv^2/4}\frac{1-\chi}{Q}\qty\big(v^2-{M'}^2)\,\omega\;\Omega^2\;\frac{\sinh((v-M')\eta)}{\sinh(\eta)}\nonumber
\end{align}
Since $1-\chi$ is compactly supported, it is easily seen that the second term is $\order{t^\infty}$, due to $\eta \neq 0$.

Neglecting it and using the same symmetries for $\beta, Q,\omega,\Omega$ under $v,M'\mapsto-v,-M'$ established in the last section, we have:
\begin{align}\label{insertedHeaviside_prior}
\gamma^J_M&=\frac{1}{\expval{1}}\sum_{M'=-J}^J\frac{\beta}{2}\int_{\mathbb{R}}\dd{v} e^{-tv^2/4}\frac{\chi}{Q}\qty\big(v^2-{M'}^2)\,\omega\;\Omega^2\;\frac{1}{\sinh(\eta)}\sum_{s=\{-,+\}}\frac{s}{2}e^{s(v-M')\eta}\nonumber\\
&=\frac{1}{\expval{1}}\sum_{M'=-J}^J\frac{\beta}{2}\int_{\mathbb{R}}\dd{v} e^{-tv^2/4}\frac{\chi}{Q}\qty\big(v^2-{M'}^2)\,\omega\;\Omega^2\;\frac{e^{(v-M')\eta}}{\sinh(\eta)}\nonumber\\
&=\frac{1}{\expval{1}}\sum_{M'=-J}^J\frac{\beta}{2}\int_{\mathbb{R}}\dd{v} e^{-tv^2/4+v\eta}\frac{\chi}{Q} P(v,M',\eta)
\end{align}
where we defined
\begin{align}\label{polynomialPbar}
e^{v\eta} P(v,M',\eta) := \qty\big(v^2-{M'}^2)\fomega(v-M',\deta/2,M')\fOmega[2](v-M',\deta/2,M')\frac{e^{(v-M')\eta}}{\sinh(\eta)}
\end{align}
Notice that $P$ would not be well defined if we were to allow $\eta=0$.
As we will see later on, the leading contributions to the expectation value are found when using that $P$ is a polynomial in $v$ and looking at the leading order coefficients ${P}_{d_J}$ and $ P_{d_J+1}$ defined by
\begin{align}
 P(v,M',\eta) =: \sum_k  P_k v^k 
\end{align}

Note that $ P(v,M',\eta)$ is at most of degree $d_J+1$. From the form of $\omega$ and $\Omega$ it transpires that its dependence on $v$ is always in the form of terms $(v\pm \deta)$ acting on $e^{v\eta}$. However, we see that $e^{-v\eta}(v+\deta)e^{v\eta}=O(v)$ while $e^{-v\eta}(v-\deta)e^{v\eta}=O(1)$, where $O$ denotes the Bachmann-Landau notation: asymptotically bounded above for $v\to \infty$. Thus, it is easy to see that $ P_{d_J+1}=0$, i.e.\ $ P(v,M',\eta)=0+O(v^{d_J})$, unless no term of the form $(v-\deta)$ appears, i.e.\ unless $M'=M$. The same argument implies for the next to leading order that $M'=M\pm1$.

Moreover, for every monomial in $(v+\deta)$ the highest order in $v$ is obtained if every $\deta$ hits $e^{v\eta}$, bringing down a further power of $v$. Consequently the next to leading order follows when $1/\sinh(\eta)$ is hit by one derivative $\deta$.
Therefore, using that $M'=M$ implies $\delta^-=\Delta^-$, $n^-=0$, $\Delta^-+n^+=J-M$, we compute:
\begin{align}
\MoveEqLeft[2] \eval{\fomega(v-M',\deta/2,M')\frac{e^{(v-M')\eta}}{\sinh(\eta)}}_{M'=M}\!\!=\!\prod_{i=1/2-\abs{M}}^{\abs{M}-1/2}\qty[\frac{v+M+\deta}{2}+i]\frac{e^{(v-M')\eta}}{\sinh(\eta)}\nonumber\\
&=\qty\Bigg(v^{2|M|}+v^{2|M|-1}\sum_{i=1/2-\abs{M}}^{\abs{M}-1/2}\qty\Big[-\frac{1}{2}\coth(\eta)+i]+O\qty\big(v^{2|M|-2}))\frac{e^{(v-M')\eta}}{\sinh(\eta)}\nonumber\\
&=\qty(v^{2|M|}-v^{2|M|-1}|M|\coth(\eta)+O\qty\big(v^{2|M|-2}))\frac{e^{(v-M')\eta}}{\sinh(\eta)}
\end{align}
\begin{align}
&\eval{\fOmega(v-M',\deta/2,M')\frac{e^{(v-M')\eta}}{\sinh(\eta)}}_{M'=M}\!\!\approx2^J\qty\Bigg(\sum_{k=0}^1(-)^k\binom{\delta^+}{k} \binom{\delta^-}{\Delta^--k}
\fomega[k](\frac{v-M-1}{2},\frac{\deta}{2}))\frac{e^{(v-M')\eta}}{\sinh(\eta)}\nonumber\\
&=2^J\qty\Bigg(\prod_{i=1}^{\Delta^-}\qty[\frac{v+M+\deta-1}{2}-J+i]-\delta^+\delta^-\qty[\frac{v-M-\deta-1}{2}]\prod_{i=2}^{\Delta^-}\qty[\frac{v+M+\deta-1}{2}-J+i])\frac{e^{(v-M')\eta}}{\sinh(\eta)}\nonumber\\
&\approx2^J\qty\Bigg(v^{\Delta^-}+v^{\Delta^--1}\qty\Bigg[\sum_{i=1}^{\Delta^-}\qty[-\frac{1}{2}\coth(\eta)-\frac{1}{2}-J+i]-\delta^+\delta^-\frac{1}{2}\qty[\coth(\eta)-1]])\frac{e^{(v-M')\eta}}{\sinh(\eta)}\nonumber\\
&=2^J\qty(v^{\Delta^-}-v^{\Delta^--1}\frac{\Delta^-+J^2-M^2}{2}\coth(\eta)
+O\qty\big(v^{\Delta^--2}))\frac{e^{(v-M')\eta}}{\sinh(\eta)}
\end{align}
where '$\approx$' denotes equality up to corrections of $O(v^{\Delta^--2})$.
Additionally, we need further subleading contributions coming from $M'=M\pm1$, which are computed similarly:
\begin{align}
\eval{\fomega(v-M',\deta/2,M')\frac{e^{(v-M')\eta}}{\sinh(\eta)}}_{M'=M\pm 1}&=\qty(v^{|2M\pm 1|}\frac{\coth(\eta)\pm1}{2}+O\qty\big(v^{|2M\pm 1|-1}))\frac{e^{(v-M')\eta}}{\sinh(\eta)}\\
\eval{\fOmega(v-M',\deta/2,M')\frac{e^{(v-M')\eta}}{\sinh(\eta)}}_{M'=M\pm 1}&=2^J\delta^-\qty(v^{\Delta^-}+O\qty\big(v^{\Delta^--1}))\frac{e^{(v-M')\eta}}{\sinh(\eta)}
\end{align}

Plugging the last four equations into (\ref{polynomialPbar}) enables us to read the coefficients
\begin{align}
P_{d_J+1}&=\frac{2^{2J}e^{-M\eta}}{\sinh(\eta)}\delta_{M,M'}\\
P_{d_J}&=\frac{2^{2J}e^{-M\eta}}{\sinh(\eta)}\qty[-\qty(J(J+1)-M^2)\coth(\eta)\,\delta_{M,M'}+\frac{(\delta^-)^2}{2\sinh(\eta)}\,\delta_{M\pm1,M'}]
\end{align}
where we used $(\coth(\eta)\pm1) = e^{\pm\eta}/\sinh(\eta)$. This finishes the proof of lemma \ref{Lemma_Coefficient}.

\subsection{Proof of Lemma \ref{Lemma_Expansion}}
We continue with (\ref{insertedHeaviside_prior}) and complete the square in the exponent w.r.t.\ $w=\sqrt{t}v-2\eta/\sqrt{t}$, i.e.\ $v=w^+$ with $w^\pm:=\pm w/\sqrt{t}+2\eta/t$. After adding the factor $\chi(w^-)=\chi(v-4\eta/t)$ to the integrand (a process found to be correct up to $\order{t^\infty}$ by an appropriate substitution) we have
\begin{align}\label{integral_to_be_taylored}
\gamma^J_M
&=\sum_{M'=-J}^J\qty(t\,e^{-t/4}\frac{\sinh(\eta)}{2\eta \sqrt{\pi}}\frac{\beta}{2})\int_{\mathbb{R}}\dd{w} e^{-w^2/4} \chi(w^+)\chi(w^-)\frac{P(w^+,M',\eta)}{Q(w^+)} \nonumber\\
&=\sum_{M'=-J}^J\qty(e^{-t/4}\frac{\sinh(\eta)}{2\eta \sqrt{\pi}}\frac{\beta}{2})\int_{\mathbb{R}}\dd{w} e^{-w^2/4} \chi(w^+)\chi(w^-) \frac{p(w,t)}{q(w,t)}
\end{align}
where we replaced $tP/Q$ by its symmetrised version w.r.t.\ $w$, using $w^+(-w)=w^-(w)$:
\begin{align}
p(w,t)&:=[{P}(w^+,M',\eta)Q(w^-)+{P}(w^-,M',\eta)Q(w^+)]\;t^{2d_J+1}/2\\
q(w,t)&:=Q(w^+)Q(w^-)\;t^{2d_J}
\end{align}
Both $p$ and $q$ are polynomials in $w$ of degree $(2d_J+1)$ and $2d_J$ respectively and symmetric in $w$. Therefore, they only contain even powers of $w/\sqrt{t}$ and, due to the monomial factor in $t$, are also polynomials in $t$ of the same respective degree.\\
To determine the power series of
\begin{align}
I(t):= \int_{\mathbb{R}}\dd{w} e^{-w^2/4} \chi(w^+)\chi(w^-) \frac{p(w,t)}{q(w,t)}
\end{align}
it can be shown that by virtue of dominated convergence we may interchange integration and limit $t\to0$ in the power series expansion (for $\eta\neq 0$):
\begin{align}
I(t)&=
\lim_{s\to 0} I(s) + t \lim_{s\to 0}(\partial_s I)(s) +\order*{t^2}\nonumber\\
&=\int_\mathbb{R}\dd{w} e^{-w^2/4}\frac{p(w,0)}{q(w,0)}+t\int_\mathbb{R}\dd{w} e^{-w^2/4}\,\partial_s\qty[\chi\qty(w^+(s))\chi\qty(w^-(s)) \frac{p(w,s)}{q(w,s)}]_{s=0}+\order*{t^2}\nonumber\\
&=\int_\mathbb{R}\dd{w} e^{-w^2/4}\left(\frac{p(w,0)}{q(w,0)}+t
\eval{\partial_s \frac{p(w,s)}{q(w,s)}}_{s=0}\right)+\order*{t^2}
\end{align}
where we neglected all terms containing the derivative of $\chi$ because it is compactly supported and the respective integral is therefore found to be $\order{t^\infty}$.\\
Denoting by $'$ the derivative in the second argument, one can convince oneself that
\begin{align*}
p(w,0)&={P}_{d_J+1}Q_{d_J}(2\eta)^{2d_J+1},\hspace{50pt}
q(w,0)=Q_{d_J}^2(2\eta)^{2d_J},\\
p'(w,0)&=-d_J{P}_{d_J+1}Q_{d_J}(2\eta)^{2d_J-1}w^2+({P}_{d_J+1}Q_{d_J-1}+{P}_{d_J}Q_{d_J})(2\eta)^{2d_J},\\
q'(w,0)&=-d_JQ^2_{d_J}(2\eta)^{2d_J-2}w^2+2Q_{d_J}Q_{d_J-1}(2\eta)^{2d_J-1}
\end{align*}
where we defined $Q(v)=:\sum_k Q_k v^k$. Using $Q_{d_J}=1,\; Q_{d_J-1}=0$ and resolving the Gaussian integrals, this leads to 
\begin{align}
I(t)=4\eta\sqrt{\pi}{P}_{d_J+1}+2\sqrt{\pi}{P}_{d_J} t+\order*{t^2}
\end{align}
Having already calculated $P_{d_J+1}$ and $P_{d_J}$, we can plug everything into (\ref{integral_to_be_taylored}). Using $\Delta^\pm=\delta^\pm \pm |M'-M|$ and $[\coth(\eta)-1/\sinh(\eta)]=\tanh(\eta/2)$, we finally get:
\begin{align}
\gamma^J_M&=\sum_{M'=-J}^Je^{-t/4}\frac{\sinh(\eta)}{4\eta\sqrt{\pi}} \qty(2^{-2J}e^{t(1-{M'}^2)/4}e^{\eta M}\frac{\Delta^+!\Delta^-!}{\delta^+!\delta^-!})\qty(4\eta\sqrt{\pi}P_{d_J+1} + 2\sqrt{\pi}P_{d_J} t +\order*{t^2}) \nonumber \\
&=\qty[1-\frac{tM^2}{4}-\frac{t}{2\eta}\qty(J(J+1)-M^2)\coth(\eta)] + \qty\bigg[\frac{t}{4\eta\sinh(\eta)}\sum_{M'=-J}^J(\delta^++1)\delta^- \,\delta_{M\pm1,M'}] +\order*{t^2} \nonumber\\
&=1-\frac{t}{4}\qty[\qty(J(J+1)-M^2)\frac{\tanh(\eta/2)}{\eta/2}+M^2] +\order*{t^2}
\end{align}
\qed

\section{Conclusion}
\label{s5}
Coherent states are an essential tool in the study of any quantum system, being able to investigate the correspondence with an emerging classical description of the system and the role of quantum fluctuations that modify it. Especially when a concrete definition of the kinematical state space of the theory is available, coherent states are the natural route to follow and might help to unravel properties of any proposal for the dynamics.\\
In this paper we repeated the construction of gauge field theory coherent states (GCS) from \cite{TW1,TW2,TW3} which are suitable for all LGTs. These GCS are labelled by classical phase space data and sharply peaked in the sense that the expectation value of any operator, corresponding to some classical function on the phase space, results in the evaluation of its classical function on said phase space data modulo higher order quantum corrections.
Moreover, we have derived the general formulas which describe the first order quantum fluctuations of these expectation values for the gauge group ${\rm SU}(2)$. Therefore we enable in principle a direct relation between novel predictions from LGT and experimental measurements.\\
Now, it would be interesting to determine these corrections for concrete models, for example already known classical solutions to ${\rm SU}(2)$ Yang-Mills theory, like those derived in \cite{Act79} and \cite{OC03}, or other systems based on this gauge group, such as Loop Quantum Gravity.\footnote{Indeed, in \cite{DL17b,LR19,HL19} this proposal is explicitly carried out for cosmological, isotropic spacetimes.}\\

Let us speculate about further applications of the analytical form of these quantum fluctuations which might help in dealing with the vast discretisation ambiguities that plague the definition of the dynamics in canonical LGT. To define the latter, one normally introduces an ultraviolet cutoff or discretisation parameter $\epsilon$ and approximates the Hamiltonian $H$ by a function $H^\epsilon$ which is solely expressed in quantities regular in the smearing parameter, such that $H=H^\epsilon+\order{\epsilon}$. While for finite $\epsilon$ a quantisation of $H^\epsilon$ is possible on its corresponding lattice Hilbert space $\mathcal{H}^\epsilon$, the continuum limit $\epsilon\to 0$ is typically problematic, e.g.\ for Yang-Mills theories the quantum Kogut-Susskind Hamiltonian $\hat H^\epsilon$ depends on inverse powers of $\epsilon$. This is the point where renormalisation techniques enter: for a family of lattices labelled by $\epsilon$ one wants to find a family $\{\mathcal{H}^\epsilon,\hat H_\star^\epsilon\}^\epsilon$ such that an inductive limit (also called direct limit \cite{KR86,Jan88}) exists to give rise to a well-defined continuum theory. The inductive limit Hilbert space $\mathcal{H}$ contains the $\mathcal{H}^\epsilon$ of all coarse lattices - loosely speaking interpretable as restrictions of the continuum theory to resolution scale $\epsilon$. An inductive limit Hamiltonian operator (once found) would generate the dynamics on the continuum Hilbert space, such that the matrix elements on states in $\mathcal H^\epsilon$ would agree with those of the Hamiltonian $\hat H_\star^\epsilon$ of finite resolution $\epsilon$. Of course, the GCS correspond to elements in some $\mathcal{H}^\epsilon$ that appear semi-classical at finite resolution $\epsilon$. And with the provided formulas the expectation value of the restriction of the continuum Hamiltonian to this state could be computed immediately.\\
However, in the light of the present formulas a different point of view also appears to be viable: Instead of considering $\mathcal{H}^\epsilon$ as restrictions of the continuum quantum field theory (QFT) to finite resolution, we might view them as auxiliary intermediate objects being interested only in the continuum theory itself, which we will interpret as the formal limit $\epsilon\to 0$. In this sense, a family of states $\{\Psi_\epsilon\}_\epsilon$ as parametrized by the lattice regulator $\epsilon$ describes a quantum state for vanishing discretisations in their limit $\epsilon\to 0$. Indeed, the GCS studied here are of this form as they are peaked over classical field content $\tilde{P}_\epsilon(S_{e_k}), \tilde{h}_\epsilon(e_k)$ (e.g.\ $\lim_{\epsilon\to 0}(\tilde h_\epsilon(e_k)-\id)/\epsilon = A_k(e_k[0])$ recovers the continuum connection). Given some observable $O$ of the gauge theory, one will discretise it to $O^\epsilon$ on a lattice $\Gamma_\epsilon$ and then quantise it as $\hat O^\epsilon$. Now, thanks to the formulas computed in (\ref{Result})-(\ref{hRRR}) it is possible to compute $\langle \Psi_\epsilon, \hat O^\epsilon \Psi_\epsilon\rangle$ for every $\epsilon>0$ where $\Psi_\epsilon$ labels a family of GCS peaked on the same continuum geometry $(E,A)$. The limit $\lim_{\epsilon\to 0} \langle \Psi_\epsilon, \hat O^\epsilon \Psi_\epsilon\rangle=O[E,A]+\hbar\; F[E,A]+\order*{\hbar^2}$ results then in the original continuum expression for $O$ evaluated on the classical field content, modulated by its \emph{continuum quantum corrections} $F$.
In total, we could therefore adapt the philosophy that - although we do not have access to the continuum QFT itself - the computed expectation values for $\epsilon\to 0 $ are speculated to carry physical relevance. This would allow for the first time to compute predictions for the quantum behaviour of a system that are not overshadowed by classical discretisation ambiguities. On the other hand, these computations could help to determine whether different discretisations $O^\epsilon$ and $\bar O^\epsilon$ would lead to different quantum corrections $F,\bar F$. In other words, we have provided a tool to check for remnants of the artificial, intermediate discretisations used to build the quantum theory.


\end{document}